\documentclass[manuscript]{aastex631}

\accepted{December 2, 2023}

\NewPageAfterKeywords
\begin{document}

\title{Dimorphos orbit determination from mutual events photometry}

\correspondingauthor{Peter Scheirich}
\email{petr.scheirich@gmail.com}

\author[0000-0001-8518-9532]{Peter Scheirich}
\affiliation{Astronomical Institute of the Czech Academy of Sciences, Fri\v{c}ova 298, CZ-25165 Ond\v{r}ejov, Czech Republic}

\author[0000-0001-8434-9776]{Petr Pravec}
\affiliation{Astronomical Institute of the Czech Academy of Sciences, Fri\v{c}ova 298, CZ-25165 Ond\v{r}ejov, Czech Republic}

\author[0000-0001-8437-1076]{Alex J. Meyer}
\affiliation{Smead Department of Aerospace Engineering Sciences, University of Colorado Boulder, Boulder, CO, USA}

\author[0000-0002-3544-298X]{Harrison F. Agrusa}
\affiliation{Université Côte d’Azur, Observatoire de la Côte d’Azur, CNRS, Laboratoire Lagrange, Nice, France}
\affiliation{Department of Astronomy, University of Maryland, College Park, MD, USA}

\author[0000-0002-0054-6850]{Derek C. Richardson}
\affiliation{Department of Astronomy, University of Maryland, College Park, MD, USA
}

\author[0000-0003-3240-6497]{Steven R. Chesley}
\affiliation{Jet Propulsion Laboratory, California Institute of Technology, CA, USA
}

\author[0000-0003-4439-7014]{Shantanu P. Naidu}
\affiliation{Jet Propulsion Laboratory, California Institute of Technology, CA, USA
}

\author[0000-0003-3091-5757]{Cristina Thomas}
\affiliation{Department of Astronomy and Planetary Science, Northern Arizona University, Flagstaff, AZ, USA}

\author[0000-0001-6765-6336]{Nicholas A. Moskovitz}
\affiliation{Lowell Observatory, 1400 West Mars Hill Road, Flagstaff, AZ 86004, USA}



\begin{abstract}

The NASA Double Asteroid Redirection Test (DART) spacecraft successfully impacted the Didymos-Dimorphos binary asteroid system on 2022 September 26 UTC.
We provide an update to its pre-impact mutual orbit and estimate the post-impact physical and orbital parameters, derived using ground-based photometric observations taken from July 2022 to February 2023.
We found that the total change of the orbital period was $-33.240 \pm 0.072$~min. (all uncertainties are 3$\sigma$).
We obtained the eccentricity of the post-impact orbit to be $0.028 \pm 0.016$ and the apsidal precession rate of $7.3 \pm 2.0$~deg./day from the impact to 2022 December 2.
The data taken later in December to February suggest that the eccentricity dropped close to zero or the orbit became chaotic approximately 70 days after the impact.
Most of the period change took place immediately after the impact but in a few weeks following the impact it was followed by additional change of 
$-27^{+19}_{-58}$ seconds or $-19 \pm 18$ seconds (the two values depend on the approach we used to describe the evolution of the orbital period after the impact -- an exponentially decreasing angular acceleration or an assumption of a constant orbital period, which changed abruptly some time after the impact, respectively).
We estimate the pre-impact Dimorphos-Didymos size ratio was $0.223 \pm 0.012$ and the post-impact is $0.202 \pm 0.018$,
which indicates a marginally significant reduction of Dimorphos' volume by ($9 \pm 9) \%$ as the result of the impact.

\end{abstract}



\section{Introduction}
\label{Introduction}

NASA's Double Asteroid Redirection Test (DART) spacecraft intentionally and autonomously collided with Dimorphos, the secondary of the Didymos binary asteroid system, on 2022 September 26 \citep{DalyEtAl2023a}.
The impact altered the orbit of Dimorphos around Didymos, reducing the orbit period by around 33 minutes,  
which was determined by two independent approaches using Earth-based photometric and radar observations \citep{ThomasEtAl2023}.

DART was NASA's first planetary defense test mission, with the goal of demonstrating the kinetic impactor mitigation technique on an asteroid.
The main benefit of targeting a binary asteroid system on a kinetic impactor mission was that it allows
the main result of the test -- the change in the mutual orbital period --
to be measured from Earth via photometric observations, assuming
that the binary system exhibits mutual events seen from Earth.
\citet{RivkinEtAl2021} discussed the factors that led to the recognition that Didymos was
the best candidate for a kinetic impactor test, and its selection as the DART
target.

Didymos was discovered in 1996 and photometric observations led to its identification as a binary system \citep{PravecEtAl2003}
and to determination of parameters of the orbit of its secondary \citep{PravecEtAl2006,ScheirichAndPravec2009}.
Radar observations of the Didymos system were also obtained in 2003, confirming that it is a binary system and characterizing its shape and orbit properties \citep{NaiduEtAl2020}.
The radar observations constrained the diameter of Didymos to about 780 m and the diameter of Dimorphos to about 150 m.

Didymos is classified as an S-type asteroid \citep{ChengEtAl2018} based
on vis-IR spectra obtained by \citet{deLeonEtAl2010}, also confirmed by \citet{DunnEtAl2013}.
Thorough spectral characterization before and after the DART impact was carried also by \citet{LinEtAl2023} and \citet{IevaEtAl2022,IevaEtAl2023}.

Following the selection of Didymos as the target of DART, a worldwide observing campaign was organised to observe the system and characterize its pre- and post-impact orbit using ground-based telescopes,
with photometric observations being a crucial aspect \citep{PravecEtAl2022,MoskovitzEtAl2023}.
These observations resulted in greatly improved knowledge of the orbit of Dimorphos around Didymos
and its orbital phase at the time of impact \citep{ScheirichAndPravec2022,NaiduEtAl2022}.

In this paper, we present an update on the pre-impact mutual orbit and results from post-impact mutual orbit modeling using the complete photometry data for mutual events in the Didymos system from 2003 to 2023.
An independent derivation of the observed mutual orbit based on an analysis of mutual event timings has been made by \citet{NaiduEtAl2023}.

For further interpretation of these results and more context on the pre- and post-impact dynamics of the Didymos system, beyond the orbit determinations, see \citet{RichardsonEtAl2023}.

A derivation of the change of the orbital period was made also by \citet{GudebskiEtAl2023}. We discuss their work in Section~\ref{Sect:Discussion}.

The Didymos system will be investigated by ESA's Hera mission from the beginning of 2027 for about half a year,
which will provide a thorough description of the post-impact state of the binary system \citep{MichelEtAl2022}.

\section{Mutual orbit model of Didymos system} \label{Model}

\subsection{Observational data} \label{ObsData}

The photometric data used in our analysis of the pre-impact orbit were obtained during six apparitions of Didymos from 2003 to 2022.
Those from apparitions from 2003 to 2021 were published in \citet{PravecEtAl2006,PravecEtAl2022}.
The data from the pre-impact apparition taken from 2022 July 2 to September 26 (part of which has been published by \citealp{ThomasEtAl2023}) are published in \citet{MoskovitzEtAl2023} 
We briefly summarize all the pre-impact data in Table~\ref{TableData21}.

\begin{table} \caption{Photometric observations of the Didymos system in the pre-impact apparitions 2003 to 2022}\label{TableData21}
\begin{center}
  \begin{tabular}{ccc}
\hline
Time span                     &  No. of runs   & Reference \\
\hline
2003-11-20.9 to 2003-12-20.3  &  17              & P06\\
2015-04-13.3 to 2015-04-14.4  &  2               & P22\\
2017-02-23.3 to 2017-05-04.3  &  13              & P22\\
2019-01-31.4 to 2019-03-11.1  &  5               & P22\\
2020-12-12.6 to 2021-03-06.3  &  15              & P22\\
2022-07-02.3 to 2022-09-26.4  &  36              & M23 \\
\hline
 \end{tabular}
\end{center}
References:
P06 \citet{PravecEtAl2006}, P22 \citet{PravecEtAl2022}, M23 \citet{MoskovitzEtAl2023}.\\
\end{table}

The data of the post-impact orbit are published in \citet{MoskovitzEtAl2023}, and we briefly summarize them in Table~\ref{TableDataPostImpact}.
In both Tables~\ref{TableData21} and~\ref{TableDataPostImpact}, we used the term {\it run} as a time series of photometric data acquired by a single facility during one night.

The data presented in Tables~\ref{TableData21} and~\ref{TableDataPostImpact} represent only the data we used in our analysis.
Since our criteria for accepting datasets was slightly more strict than those used by \citet{MoskovitzEtAl2023}, our data are a higher-quality subset,
containing 15\% less runs, of that reported by \citet{MoskovitzEtAl2023}.

\begin{table} \caption{Photometric observations of the Didymos system in the post-impact apparition 2022-2023}\label{TableDataPostImpact}
\begin{center}
  \begin{tabular}{cccc}
\hline
Lunation & Time span                &   Days from impact  &  No. of runs    \\
\hline
L0  & 2022-09-28.2 to 2022-10-10.3  &   1.1 to 13.4       &               47              \\     
L1  & 2022-10-24.4 to 2022-11-02.5  &   27.4 to 36.5      &               20               \\
L2  & 2022-11-17.4 to 2022-12-02.3  &   51.3 to 66.5      &               48              \\
L3  & 2022-12-14.4 to 2022-12-31.3  &   78.4 to 95.4      &               25               \\
L4  & 2023-01-11.2 to 2023-01-30.1  &   106.1 to 125.3    &               13              \\
L5  & 2023-02-11.2 to 2023-02-21.3  &   137.1 to 147.4    &                4              \\
\hline
 \end{tabular}
\end{center}
\end{table}

The data were analysed using the standard technique described in \citet{PravecEtAl2006,PravecEtAl2022}.
Briefly, by fitting a two-period Fourier series to data points taken outside mutual (occultation or eclipse) events, the rotational lightcurves of the
primary and the secondary, which are additive in flux units, are separated.
Subtracting the rotational lightcurve of the primary from the data, a long-period (orbital) lightcurve component containing the mutual events and the secondary rotation lightcurve is
obtained, which is then used for subsequent numerical modeling. We refer the reader to \citet{PravecEtAl2022} for details of the lightcurve decomposition method.

\subsection{Numerical model} \label{NumModel}

We constructed a model of the Didymos-Dimorphos system adapting the technique of \citet{ScheirichAndPravec2009} which was further developed by \citet{ScheirichEtAl2015,ScheirichEtAl2021} and used also by \citet{ScheirichAndPravec2022}
and \citet{ThomasEtAl2023}.
We outline the basic points of the method below, but we refer the reader to the above references for details of the technique.

For the pre-impact orbit, we applied the method used by \citet{ScheirichAndPravec2022} (hereafter referred
to as SP22) on the complete pre-impact dataset (Table~\ref{TableData21}).
For the post-impact orbit, we made a few modifications, which we describe in Section~\ref{PostImpactParams}.
The method of SP22 is described in following.

The binary asteroid components were represented with an oblate ellipsoid for the primary and an oblate ellipsoid or a sphere (in the case of the post-impact orbit, see Section~\ref{PostImpactParams}) for the secondary.
(The shapes of the bodies were used to calculate the light flux of the system only.
The orbital characteristics of the system were treated as free parameters; they were not tied to the dynamics that would come from the non-spherical shapes.)
% DONE: Harrison: (If I understand correctly) I would be clear that just the magnitude of the system is computed assuming oblate ellipsoids / spheres. Since you mention the orbit is assumed to be Keplerian, so they may be confused about the use of nonspherical shapes.
For the pre-impact orbit, we choose a circular orbit for simplicity, as the upper limit on the eccentricity is low (see below).
The circular pre-impact orbit is also consistent with the observed post-impact eccentricity of $\sim 0.03$ (see Section~\ref{PostImpactParams}), as noted by \citet{MeyerEtAl2023} (also mentioned in \citealp{RichardsonEtAl2023}).
% DONE: Alex: The post-impact eccentricity of ~0.03 is also consistent with a circular pre-impact orbit (see Meyer et al 2023, also mentioned in Richardson et al 2023)
The motion was assumed to be Keplerian, but we allowed for a quadratic drift in the mean anomaly, which results from the combined effects of BYORP and tidal dissipation or from differential Yarkovsky force (SP22).
% DONE: Harrison: … mean anomaly, which results from the combined effects of BYORP and tidal dissipation.
The spin axis of the primary was assumed to be normal to the mutual orbital plane of the components.
The shapes were approximated with 1016 and 252 triangular facets for the primary
and the secondary, respectively.
The components were assumed to have the same albedo and their surfaces uniform in albedo and scattering law.
The brightness of the system as seen by the observer was computed as a sum of contributions from all visible facets using a
ray-tracing code that checks which facets are occulted by or are in shadow from the other body.
A combination of Lommel-Seeliger and Lambert scattering
laws was used (see, e.g., \citealp{KaasalainenEtAl2002}).

The quadratic drift in the mean anomaly, $\Delta M_d$, was fitted as an independent parameter on the pre-impact data.
It is the coefficient in the second term of the expansion of the time-variable mean anomaly:
\begin{equation}
M (t) = M (t_0) + n (t-t_0) + \Delta M_d (t - t_0)^2, \label{dMd1}
\end{equation}
where
\begin{equation}
\Delta M_d = \frac{1}{2} \dot{n}, \label{dMd2}
\end{equation}
where $n$ is the mean motion, $t$ is the time, and $t_0$ is the epoch.
$\Delta M_d$ was stepped from $-10$ to $+10$~deg/yr$^2$ with a step of 0.01~deg/yr$^2$,
and all other parameters were fitted at each step.

Across all pre-impact observations, we found a unique solution for the system parameters, see Table~\ref{tablePreImpact}.
We describe these parameters in Section~\ref{PreImpactParams}.

For the post-impact observations, we developed two alternative modifications of the model in order to describe specific post-impact features seen in the data.
For both modifications, we found unique solution for their parameters, both describing the data equally well.
We describe and discuss the modifications and their parameters in Section~\ref{PostImpactParams}.

We estimated uncertainties of the fitted parameters using two techniques.
The uncertainty of the orbital pole (the pole is strongly determined by the shapes of the mutual events) was
estimated using the procedure described in \citet{ScheirichAndPravec2009}.
The uncertainties of the rest of the parameters, which are determined primarily by the timings of the events,
were estimated using the method described in \citet{ScheirichEtAl2021},
which we outline below.

The residuals of the model fitted to the observational data do not obey the Gaussian statistics because of systematic errors resulting from model simplifications.
In particular, the residuals of nearby measurements appear correlated. To eliminate the effect we adopted the following strategy based on the $\chi^2$ test.

We choose a correlation time $d$ and for each data point ($i$) we calculated how many other data points, $K_i$, are within $\pm d/2$ from the given point.
We then applied a weight of $1/K_i$ to the given data point in the $\chi^2$ sum.
We also calculated an effective number of data points as $N_{\rm eff} = \sum^N_{i=1} 1/K_i$, where $N$ is the total number of data points.
For normalized $\chi^2$ we then have $\chi^2 = 1/(N_{\rm eff}-M) \sum^N_{i=1} (O-C)_i^2/(\sigma_i^2 K_i)$, where $M$ is the number of fitted parameters of the model and $\sigma_i$ is a standard deviation of the $i$th point.
As the residuals are predominated by model rather than observational uncertainties, we assign each data point the same standard deviation $\sigma_i = \sigma$, where $\sigma$
is the RMS residual (root mean square of observed magnitudes, $O$, minus the values calculated from the model, $C$) of the best fit solution.
An illustration of the weights $1/K_i$ determination is shown in Figure~2 of SP22.

The procedure described above is equivalent to reducing the number of data points to one in each time interval with the length $d$ (i.e., to reducing the total number of points to $N_{\rm eff}$)
and assigning $(O-C)^2$ of this point to be a mean of $(O_i-C_i)^2$ of all the points within the interval.
However, our approach has the advantage that it does not depend on a particular realization of dividing the observing time span into the intervals of length $d$.

We choose the correlation time $d$ to be equal to 1/2 of the mean duration of a decreasing/increasing branch\footnote{We define a {\it branch} as a part of the mutual event in the lightcurve,
where the brightness of the system is rapidly decreasing or increasing, i.e., the time period during which the eclipsed/occulted body is immersing into or emerging from the shadow of (for a mutual eclipse events),
or is disappearing behind or reappearing from behind
the other body (for a mutual occultation event).}
of the secondary mutual event,
i.e., the mean time between the first and the second or between the third and the fourth contact between the limbs or terminators of the two bodies in their projection onto the sky.
For the observed events in Didymos, it is $d=0.14$~h.
(We also tested $d$ to be twice as long, i.e., equal to the full mean duration of the secondary event branch, but we found it to be inadequate as the longer correlation time resulted in
a substantial loss of information
by deweighting the datapoints too much.)

We note that the mutual orbit model fit is sensitive only to data points covering mutual events and their closest neighborhood.
Therefore we limited the above analysis only to such data points; points further outside the events were not used, because they do not effectively contribute to the determination of the mutual orbit.

Upon stepping a given parameter on a suitable interval (while the other parameters fitted) and computing the normalized $\chi^2$ for each step,
we determined 3-$\sigma$ uncertainty of the given parameter as an interval in which $\chi^2$ is below the p-value of the $\chi^2$ test, corresponding to the
probability that the $\chi^2$ exceeds a particular value only by chance equal to 0.27\%.

\newpage
\section{Parameters of Didymos-Dimorphos system}\label{Params}

\subsection{Pre-impact orbit} \label{PreImpactParams}

\begin{table} \caption{Parameters of the Didymos-Dimorphos pre-impact model. The indices 1 and 2 refer to Didymos and Dimorphos, respectively.} \label{tablePreImpact}
\begin{center}
\makebox[\textwidth]{
\begin{tabular}{lccc}
\hline
Parameter                              &    &  Value                            &  Unc.     \\
\hline
\multicolumn{4}{l}{Secondary:}  \\
Cross-section equiv. diam. ratio & $D_{2,C}/D_{1,C}$                  & $0.220 \pm 0.004^a$               & 1$\sigma$ \\
Volume equivalent diameter ratio & $D_{2,V}/D_{1,V}$                  & $0.223 \pm 0.004$               & 1$\sigma$ \\
\hline
\multicolumn{4}{l}{Mutual orbit:} \\
Sem. axis / primary equat. diam. & $a/(A_1 B_1)^{1/2}$                        & $1.49  \pm 0.13 $  & 3$\sigma$ \\
Ecl. longitude of orbital pole & $L_P$ (${}^{\circ}$)                  & $310.0 \pm 15.0^b$                & 3$\sigma$  \\
Ecl. latitude of orbital pole & $B_P$ (${}^{\circ}$)                 & $-80.4 \pm 1.9^b$                   & 3$\sigma$  \\
Drift in mean anomaly & $\Delta M_d$ (deg/yr$^2$)                  & $0.13 \pm 0.10$                   & 3$\sigma$ \\
Mean motion rate & $\dot{n}$ (rad/s$^2$)                           & $4.56 \pm 3.51 \times 10^{-18}$   & 3$\sigma$ \\
Pre-impact orbital period at $t_{\rm imp}$ & $P_{\rm orb}^{\rm pre}$ (h)        & $11.921493 \pm 0.000091$      & 3$\sigma$ \\
Eccentricity & $e$                                        & $\le 0.03$                        & 3$\sigma$  \\
Epoch of impact & $t_{\rm imp}$  & \multicolumn{2}{l}{JD 2459849.46875 (geocentric UTC)}\\
\hline
 \end{tabular}}
\end{center}
$^a$ The ratio is for the average observed aspect of 14.0$^{\circ}$. See text for details.
$^b$ For the actual shape of the uncertainty area, see Figure~\ref{LB_polar}. Semiaxes of the area are 2.0 $\times$ 2.5$^{\circ}$.
\end{table}

Using the data taken from 2022 July 2 to September 26, added to the 2003-2021 dataset from SP22, we applied the method of SP22,
modeling the system with a circular orbit with the quadratic drift of the mean anomaly.
We obtained updated values of the pre-impact mutual orbit parameters, which are given in Table~\ref{tablePreImpact} together with their uncertianties.
The indices 1 and 2 refer to Didymos and Dimorphos, respectively.

Unlike in SP22, we now used oblate ellipsoids as the approximation of the shapes of the components, with their axes ratios derived from best-fit ellipsoids of shape models obtained from DART data.
\citet{BarnouinEtAl2023} gives the principal axes of the best-fit ellipsoid of Didymos to be $A_1 = 819 \pm 14$~m, $B_1 = 796 \pm 14$~m and $C_1 = 590 \pm 14$~m.
\citet{DalyEtAl2023b} gives the principal axes of the best-fit ellipsoid of Dimorphos to be $A_2 = 173 \pm 1$~m, $B_2 = 170 \pm 4$~m and $C_2 = 114\pm 1$~m.
%Using those values, we derived $A_1/C_1 = (a_1 + b_1)/(2 c_1) = 1.369$ as the nominal axes ratio of the primary and $A_2/C_2 = (a_2 + b_2)/(2 c_2) = 1.504$ as the nominal axes ratio of the secondary.
Using those values, we derived $(A_1 B_1)^{1/2}/C_1 = 1.369$ as the nominal axes ratio of the oblate primary model and $(A_2 B_2)^{1/2}/C_2 = 1.504$ as the nominal axes ratio of the oblate secondary model.

% DONE: Olivier: Terik is preparing a new paper that will discuss the final Dimorphos model in the DART special issue PSJ (But you can reference his ACM abstract for now)
% E. Palmer is also preparing a new paper that will discuss the final Didymos model in the DART special issue PSJ (But you can reference his ACM abstract for now)
% These models are also just delivered this week to PDS - and there will be a PDS reference too - by the time you publish this. You can probably reference dart shape model bundle at PDS . I'll give you  a PDS LID/DOI once I get one.

% DONE: Harrison: Terik is going to submit a paper with an updated shape for Dimorphos, So you could reference Daly+ 2023, submitted for now . (will still need a separate reference for Didymos I think)

Despite the size ratio of the two components could be computed from the above axes estimates, we fitted it as an independent parameter.
This is because the native parameter of our model is the ratio between
$D_{1,C}$ and $D_{2,C}$, the mean (rotationally averaged) cross-section equivalent diameters (i.e., the diameter of a sphere with the same cross-section) of the primary and secondary, respectively,
at the mean aspect of observed total secondary events (see below).
$D_{1,C}$ and $D_{2,C}$ are determined by the actual shapes of the components as well as by the specific geometry of the binary system with respect to the observer and the Sun during the observations.
%As the best-fit ellipsoids from \citet{PalmerEtAl2023} and \citet{DalyEtAl2023b} were derived from a single geometry (from the direction of the DART approach), $D_{2,C}/D_{1,C}$ computed from them may not match that derived from our data.

To quantify the mean aspect we used an asterocentric latitude of the Phase Angle Bisector (PAB), which is the mean direction between the heliocentric and geocentric directions to the asteroid.
As discussed in \citet{HarrisEtAl1984}, this is an approximation for the effective viewing direction of an asteroid observed at the non-zero solar phase.
The average absolute value of the asterocentric latitude of the PAB for the observed total events across the pre-impact data was $14.0^{\circ}$. (We computed the latitude of the PAB using the nominal pole of the mutual orbit and
assuming that the spin poles of both components are the same as the orbit pole.)

In order to quantitatively compare the size of Dimorphos before and after the impact (see Section~\ref{PostImpactParams}),
we also computed $D_{2,V}/D_{1,V}$, a ratio of volume equivalent diameters (i.e., the diameter of a sphere with the same volume) of the components from $D_{2,C}/D_{1,C}$ and axes ratios of their models.

$a/(A_1 B_1)^{1/2}$ is the ratio between the semimajor axis of the system and mean equatorial diameter of the primary.
$L_P, B_P$ are the ecliptic coordinates of the orbital pole in the equinox J2000,
$P_{\rm orb}^{\rm pre}$ is the pre-impact orbital period at the nominal time of the DART impact and $\Delta M_d$ is the quadratic drift in the mean anomaly.
We also give the time derivative of the mean motion $\dot{n}$, calculated from $\Delta M_d$.
$e$ is the orbit eccentricity (only its upper limit was obtained before the impact, see below).

\begin{figure}[ht!]
\plotone{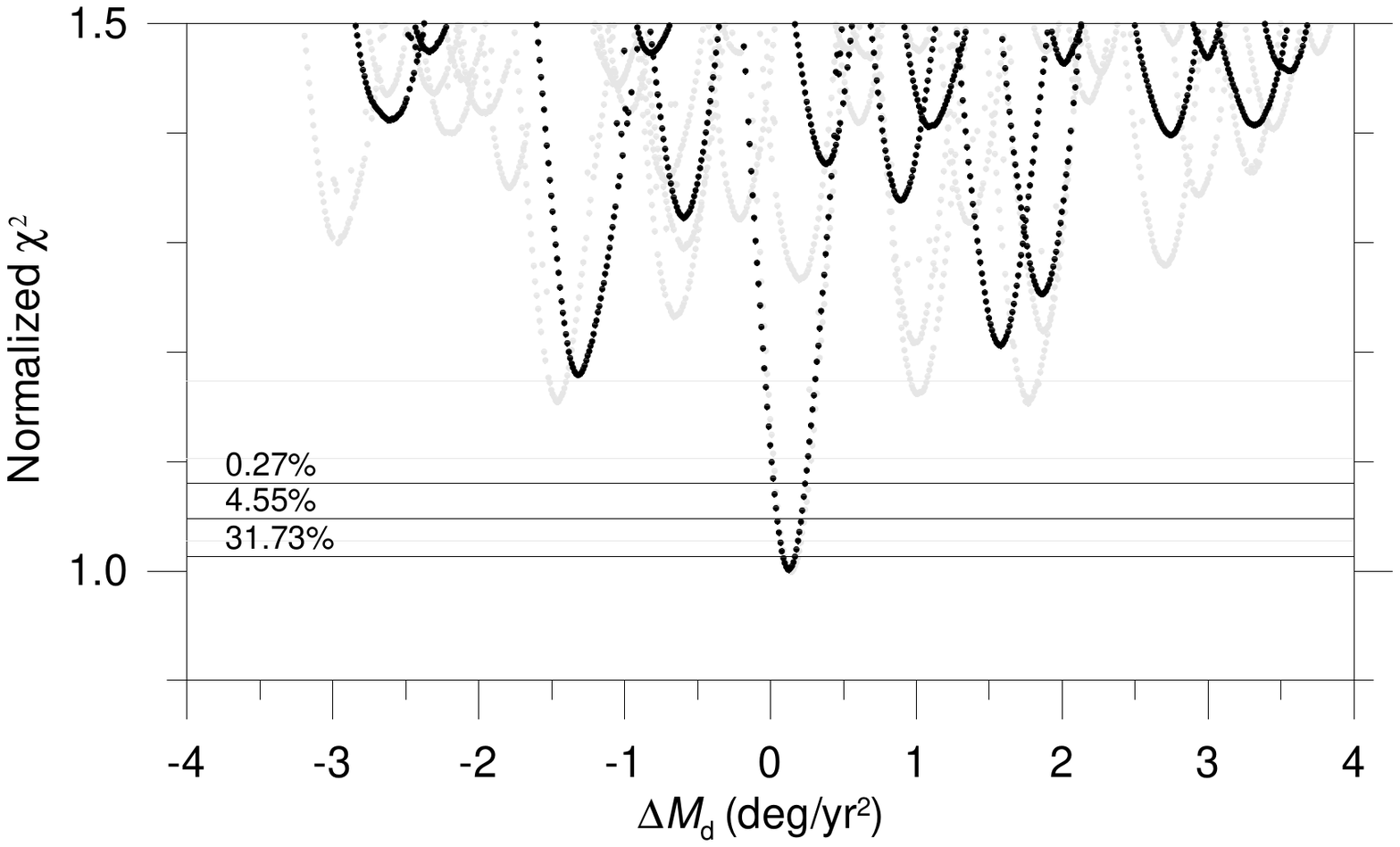}
\caption{The normalized $\chi^2$ vs. $\Delta M_d$ for solutions of the pre-impact orbit model presented in Section~\ref{NumModel}.
The three horizontal lines give the p-values -- the probabilities that the $\chi^2$ exceeds a particular value only by chance, corresponding to 1-, 2- and $3\sigma$ interval of the $\chi^2$ distribution.
The light-gray dots and lines are taken from SP22.
The best-fit value has $\Delta M_d = 0.13$~deg/yr$^2$.
\label{Chi2Norm_vs_DMd}}
\end{figure}

\begin{figure}[ht!]
\plotone{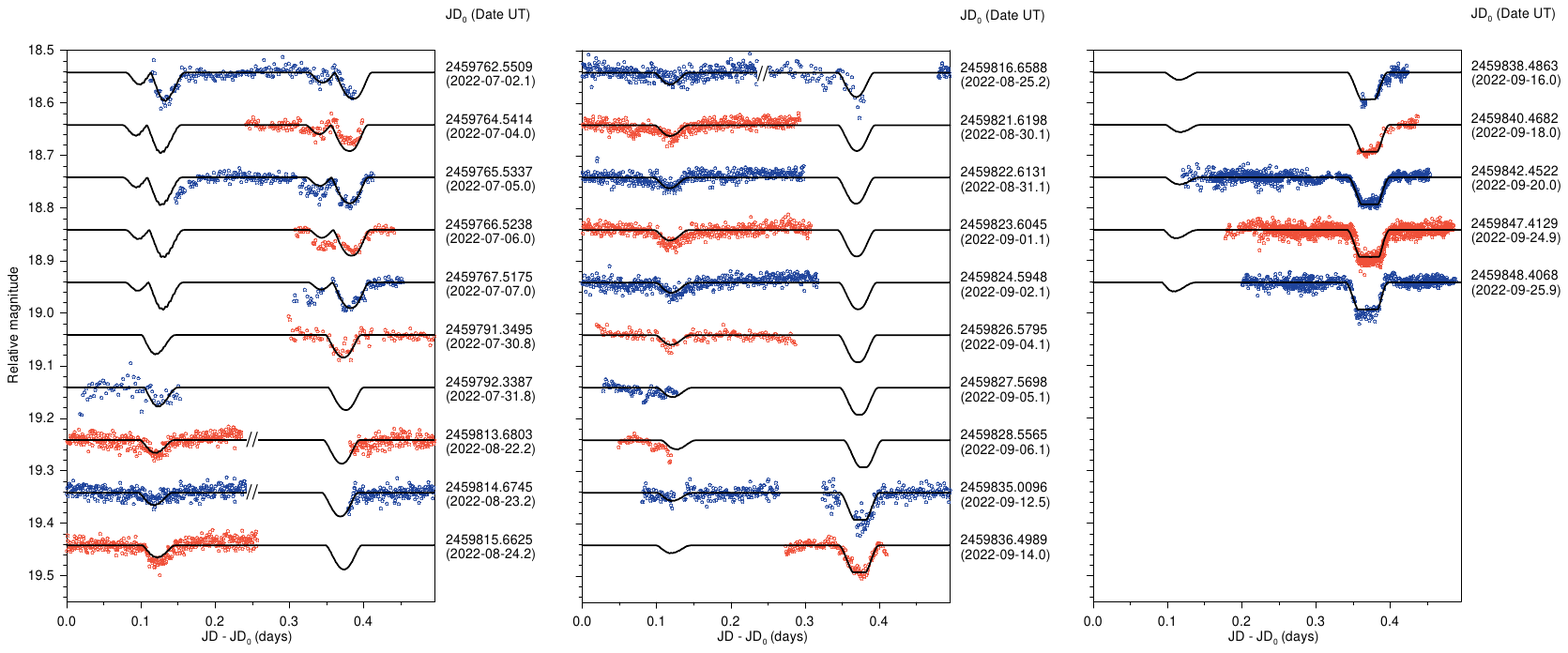}
\caption{The pre-impact orbital lightcurve component of the Didymos system in 2022. The observed data are marked as points.
(To avoid confusion between different data sets, different colors are used alternating on each plot.)
The solid curve represents the synthetic lightcurve for the best-fit pre-impact solution.
The primary and secondary events (the terms refer to which of the two bodies is occulted or eclipsed) are always shown on the left and right sides of the plots, respectively.
In some cases, the observations of a secondary event preceded that of a primary event (i.e., their order in the dataset is inverse of that shown on the plot).
In order to save space in the plot, we present these events in reverse order to how they were observed.
They are separated by the "//" symbol in the plot and one orbital period (0.496~d) is to be subtracted from the x coordinate of data points to the right from this separator.
\label{Didymos_synth_preimpact}}
\end{figure}

Plot of the normalized $\chi^2$ vs $\Delta M_d$ is shown in Figure~\ref{Chi2Norm_vs_DMd}.
The long-period (orbital) lightcurve component data from 2022 together with the synthetic
lightcurve of the best-fit solution are presented in Figure~\ref{Didymos_synth_preimpact}.
% DONE: Harrison: can you make this a single figure with three subfigures?
The synthetic curve of the best-fit solution looks indistinguishable from the synthetic curve from SP22 fitted to the data from 2003 -- 2021.
To save space, we therefore include only plots for the pre-impact 2022 apparition in this paper.

\begin{figure}[ht!]
\plotone{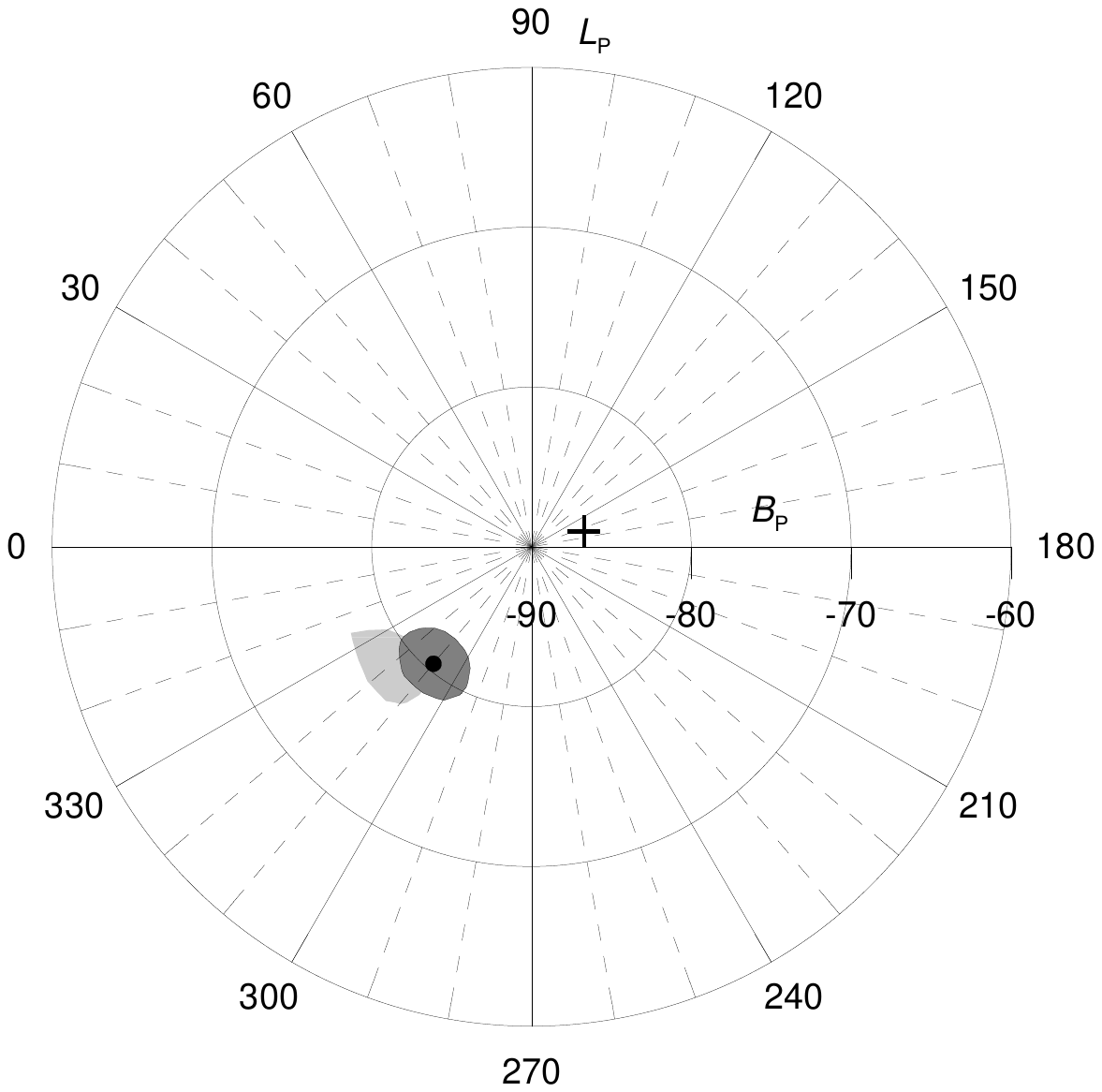}
\caption{Area of admissible poles for the pre-impact mutual orbit of Didymos in ecliptic coordinates (dark grey area).
The dot is the nominal solution given in Table~\ref{tablePreImpact}.
This area corresponds to $3\sigma$ confidence level.
The south pole of the current asteroid's heliocentric orbit is marked with the cross.
For comparison, the area of admissible poles from SP22 is shown as light gray area.
\label{LB_polar}}
\end{figure}

The uncertainty area of the orbital pole is shown in Figure~\ref{LB_polar}.
We note that the difference between the orbital pole determined in this work and that derived in SP22
is mainly due to the fact that we now used the ellipsoidal oblate model for Didymos derived from DART data.
The axes ratio used in SP22 was systematically off,
as the axes ratio we used here (1.369) is close to the 3-$\sigma$ upper limit of Didymos axes ratio assumed in SP22.
The difference in the orbital pole also explains the different value of $D_{2,C}/D_{1,C}$ we obtained with respect to the value from SP22 ($0.220 \pm 0.004$ vs. $0.217 \pm 0.004$),
as the mean aspect is slightly different.

No significant change of the solution was found for the polar flattening of the primary in a range from 1.21 to 1.55 and the polar flattening of the secondary in a range from 1.40 to 1.61.
We calculated these limits using the 1-$\sigma$ uncertainties from \citep{BarnouinEtAl2023} and \citep{DalyEtAl2023b} (see above), which we multiplied by 3 to get 3-$\sigma$ uncertainties.

\citet{ScheirichAndPravec2009} estimated the upper limit on the eccentricity using the data from the 2003 apparition.
SP22 checked that the 3-$\sigma$ upper limit of 0.03 is consistent with the data taken from 2015 to 2021.
In this work, we checked that it is consistent with the data taken from 2022 July 2 to September 26 as well, but those later data do not possess the characteristics necessary to use them for constraining the eccentricity more.
Those characteristics include sufficient quality required to resolve offsets in event timings with respect to the solution with circular orbit,
and time coverage required to obtain a robust estimate on the apsidal precession. Thus, the upper limit on the eccentricity by \citet{ScheirichAndPravec2009} is still valid.

\newpage
\subsection{Post-impact orbit} \label{PostImpactParams}

For modeling the post-impact data, we fixed the orbital pole of the mutual orbit on its pre-impact value.
% DONE: Derek:  I'm confused by this — the post-impact semimajor axis should be smaller than the pre-impact value, right?
% DONE: Harrison: I agree, I think a referee would find this odd no? The pole should be fixed to the same value, but the semimajor axis should definitely be smaller. Would the orbital solution change much if you fixed the semimajor axis to a different value?
% DONE: Cristina: It states that you set the semi-major axis to be the pre-impact value. Won't that effectively set the orbital period to be the pre-impact period?
We used an oblate ellipsoid as the approximation of the shape of Didymos, with its axes ratios the same as used for the pre-impact orbit modeling (see Section~\ref{PreImpactParams}).

We chose to approximate the post-impact Dimorphos with a sphere for following reasons.
As shown by \citet{AgrusaEtAl2021} and \citet{MeyerEtAl2023}, it is possible that Dimorphos entered a tumbling state a few tens of days after the impact due to resonances among system's natural frequencies.
When that happened, the orbital eccentricity dropped in just a few days as a result of the exchange of angular momentum between the mutual orbit and the secondary's rotation state.
When tumbling occured, the eccentricity also varied chaotically in time.
% DONE: Alex: Can mention the decrease is relatively rapid and large, taking only a few days.
% DONE: Harrison: ...would drop as a result of the exchange of angular momentum between the mutual orbit and the secondary’s rotation state. (Also, we should mention here that if tumbling occurs, the eccentricity does indeed drop. But it also varies chaotically in time)
As we do see an indication of the eccentricity drop in the data (see below), we assume that Dimorphos began tumbling in a course of the post-impact apparition.
With the chaotically spinning Dimorphos, we choose to approximate Dimorphos with a sphere for simplicity.

\begin{table} \caption{Parameters of the Didymos-Dimorphos post-impact model. The indices 1 and 2 refer to Didymos and Dimorphos, respectively.} \label{tablePostImpact}
\begin{center}
\makebox[\textwidth]{
\begin{tabular}{lccc}
\hline
Parameter                              &    &  Value                            &  Unc.     \\
\hline
\multicolumn{4}{l}{Secondary:}  \\
Cross-section equiv. diam. ratio & $D_{2,C}/D_{1,C}$                  & $0.209 \pm 0.018^a$               & 3$\sigma$ \\
Volume equiv. diam. ratio & $D_{2,V}/D_{1,V}$                         & $0.202 \pm 0.018$               & 3$\sigma$ \\
\hline
\multicolumn{4}{l}{Mutual orbit:} \\
Sem. axis / primary equat. diam. & $a/(A_1 B_1)^{1/2}$                               & $1.46  \pm 0.11 $               & 3$\sigma$ \\
Ecl. longitude of orbital pole  & $L_P$ (${}^{\circ}$)                 & $310.0 \pm 15.0 ^b$                & 3$\sigma$  \\
Ecl. latitude of orbital pole  & $B_P$ (${}^{\circ}$)                 & $-80.4 \pm 1.9 ^b$                   & 3$\sigma$  \\
Final post-impact orbital period $(t=\infty)$& $P_{\rm orb}^{\rm post}$ (h)      & $11.3675 \pm 0.0012$        & 3$\sigma$ \\
Total period change            & $\Delta P (t=\infty) $ (min.)                   & $-33.240 \pm 0.072$                    & 3$\sigma$ \\
Eccentricity in L0 to L2 & $e_{0-2}$                                & $ 0.028 \pm 0.016$                        & 3$\sigma$  \\
%Eccentricity in L0 & $e_0$                                   & $ 0.042 \pm 0.021$                        & 3$\sigma$  \\
Apsidal precession rate & ${\rm d}\varpi / {\rm d} t$ (deg./day)       & $ 7.3 \pm 2.0$                          & 3$\sigma$  \\
\multicolumn{4}{l}{\quad Solution with the angular acceleration model:} \\
Angular acceleration & $A$ (deg/d$^2$)               & $ (7.3 \pm 5.0)/\tau^2$                     & 3$\sigma$  \\
Angular acceleration timescale & $\tau$ (d)                 & $ 14.3 \pm 6.5$                        & 3$\sigma$  \\
Orbital period at $t=t_{\rm imp}$ & $P_{\rm orb}^{\rm imp}$ (h) & $11.3751^{+0.0160}_{-0.0052}$ & 3$\sigma$  \\
Period change at $t=t_{\rm imp}$  & $\Delta P (t=t_{\rm imp})$ (min.) & $-32.78^{+0.96}_{-0.31}$ & 3$\sigma$  \\
\multicolumn{4}{l}{\quad Solution with orbital period modeled as step function:} \\
Mean orbital period before the period drop & $P_{\rm orb}^0$ (h)    & $11.3729 \pm 0.0050$                      & 3$\sigma$  \\
Time of the period drop (after the impact) & $\Delta T$ (d) & $15.5^{+12.0}_{-7.0}$          & 3$\sigma$  \\
Period change at $t=t_{\rm imp}$  & $\Delta P (t=t_{\rm imp})$ (min.) & $-32.92 \pm 0.30$ & 3$\sigma$  \\
\hline
 \end{tabular}}
\end{center}
$^a$ The ratio is for the average observed aspect of 9.6$^{\circ}$. See text for details.
$^b$ Fixed on the pre-impact value.

\end{table}

An independent fit to the post-impact data (without those from the L0 lunation, which have the apparent depths of mutual events reduced due to strong contamination by the ejecta) gives $D_{2,C}/D_{1,C} = 0.209 \pm 0.018$ (3$\sigma$).
The average absolute value of the asterocentric latitude of the PAB for the observed total events across the post-impact data was $9.6^{\circ}$.
% Pre-impact:
%   D2c/D1c = 0.220 +/- 0.004 (1-sigma). Mean PAB_beta = 14°.
%   Scheirich and Pravec 2022 davaji: D2c/D1c = 0.217 +/- 0.004, mean PAB_beta = 9.7°.
%   Pro zplostely primar a kulovy sekundar plati, ze cim je PAB_beta mensi, tim mensi se jevi D1c, a proto D2c/D1c je tim vetsi.
%   Pokud je ale sekundar zplostelejsi nez primar, pak je ta zavislost obracena: cim je PAB_beta mensi, tim mensi je D2c/D1c. Coz je presne to, co u pre-impact vysledku pozorujeme.
%
% Post-impact D2c/D1c = 0.209 +/- 0.018 (3-sigma). Mean PAB_beta = 9.6°.
%   Ve srovnani s pre-impact je D2c/D1c mensi, a to i ve srovnani s hodnotou 0.217 +/- 0.004 pro mean PAB_beta = 9.7°.
%   Ve srovnani s mean PAB_beta = 14° by D2c/D1c melo vyjit mensi: pokud je sekundar zplostelejsi nez primar, pak cim je PAB_beta mensi, tim je D2c/D1c mensi. To ale plati POUZE pro pripad, ze poly primaru i sekundaru jsou koplanarni.
%   Pokud sekundar tumbluje, tak by se naopak mel zdat byt efektivne vetsi.
%   Pro zplostely primar stale plati, ze cim je PAB_beta mensi, tim mensi se jevi D1c, a tudiz by D2c/D1c melo byt vetsi.
%   Oba tyto body vyse by mely prispet k tomu, ze uvidime D2c/D1c vetsi, ale mi ho ve skutecnosti vidime mensi.

The cross-section of an oblate ellipsoid is lowest when observed equator-on and highest when observed pole-on.
As the mean asterocentric latitude of the PAB in the post-impact data was lower than in the pre-impact data ($9.6^{\circ}$ vs. $14.0^{\circ}$), the effective post-impact $D_{1,C}$ was lower than it was pre-impact.
Also, the mean cross-section of a randomly oriented (tumbling) oblate shape is higher than that of the same shape observed equator-on.
Therefore, the effective $D_{2,C}$ in the post-impact data would be higher than it was pre-impact in a case Dimorphos' shape did not change.
These two things would therefore independently lead to the post-impact $D_{2,C}/D_{1,C}$ being greater than the pre-impact, if the size and shape of Dimorphos had not change.
Despite this expectation, the observed post-impact value is lower than the pre-impact value at approximately 2-$\sigma$ level, which indicates that the size of Dimorphos has decreased as a result of the impact.
Because $D_{2,C}/D_{1,C}$ is dependent on the observing geometry for the reasons above, we also calculated the ratio of volume equivalent diameters $D_{2,V}/D_{1,V}$.
The difference between its post- and pre-impact values ($0.202 \pm 0.006$ vs. $0.223 \pm 0.004$) indicates reduction of Dimorphos' volume by ($9 \pm 3) \%$ (1-$\sigma$ uncertainties).

The spherical approximation of Dimorphos prevent us from fitting the shape-induced rotational lightcurves.
The scope of this paper is limited to obtaining the parameters of the mutual orbit.
We note that we did see the Dimorphos' rotational lightcurves in the best data that we took in December 2022 and January 2023,
suggesting some degree of tumbling, and we will study this topic statistically in another paper (\citealp{PravecEtAl2023}).

% DONE: Derek:  I'm also confused by this — doesn't this prevent fitting the shape-induced lightcurve amplitude? Or is the idea we're just fitting the orbit here (mutual events) and not all the lightcurve details?
% PP: Yes, in this work Petr S. fits the mutual events but not the Dimorphos's rotational lightcurves that we observed on the 7 epochs in December and January (see my ACM poster).
% We are going to study the Dimorphos's lightcurves (suggesting some degree of tumbling) statistically with Alex, but our paper Pravec, Meyer, Scheirich, et al. on the post-impact Dimorphos's rotational lightcurves will be done later (not sure that we meet the September 25 deadline for the PSJ focus issue, we may submit it somewhere else later in the year).  Sorry that the study takes time, but there are a number of points that we need to solve with Alex so that to do the statistical study right and obtain (hopefully) robust constraints on the Dimorphos's post-impact shape and spin state.

\begin{figure}[ht!]
\plotone{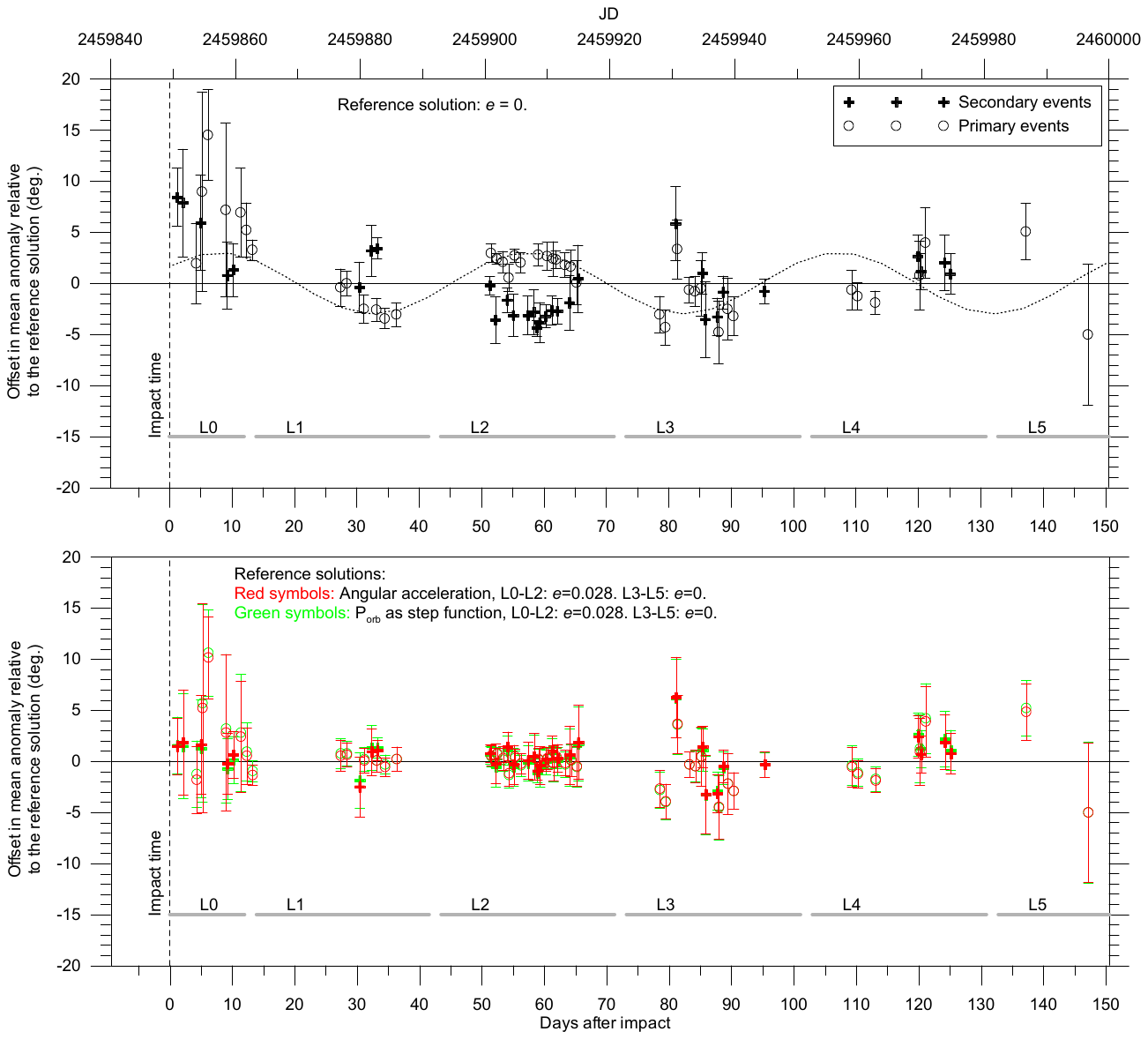}
\caption{Offsets in the mean anomaly of the observed mutual events in the post-impact apparition with respect to three reference solutions.
Top panel: the offsets are shown with respect to the reference solution that has a constant orbital period and zero eccentricity.
The dotted line is an arbitrary sine function with a period of 50 days, plotted there in order to highlight the periodicity of the systematic shifts between the primary and the secondary events during L0, L1 and L2 lunations.
Bottom panel: the offsets are shown with respect to the two nominal solutions presented in Table~\ref{tablePostImpact}, one having the exponentially decreasing angular acceleration (red symbols) and the second having the orbital period described as a step function (green symbols).
See text for details.
The gray bars at the bottom of each panel indicate spans of the individual lunations (L0 to L5; the full moon occurred in each gap between the bars).
\label{Fig:offsets_mean_anomaly}}
\end{figure}

In order to check for the eccentricity and/or its evolution, or for an evolution of the orbital period, we used the following approach.
We first fit the post-impact data using a circular orbit with a constant mean orbital period as a reference solution.
We then constructed Figure~\ref{Fig:offsets_mean_anomaly} (top panel) showing offsets in mean anomaly with respect to the reference solution for every observed mutual event
(here we used only completely covered events, i.e., that have both the decreasing and the increasing branches covered by the data).
It was computed as follows. We generated a synthetic lightcurve using the model parameters from the reference solution.
Then, for each mutual event, we fitted the mean anomaly of the model in order to obtain the best match between the synthetic lightcurve and the observed data.
The values on the vertical axis of Figure~\ref{Fig:offsets_mean_anomaly} are differences between the mean anomaly of the reference model with circular orbit and the fitted one that was shifted in mean anomaly.
In other words, the calculated differences correspond to the amount Dimorphos is ahead or behind in time with respect to an ephemeris from the reference solution of the orbit, expressed in degrees of the orbital phase
(note that 1 deg. corresponds to $P_{\rm orb} / 360 \simeq 1.9$~min.)

For each event, we computed also a standard deviation of the offset using the procedure described in Section~\ref{NumModel}, but with $\chi^2$ computed only from the data points in the vicinity of the given mutual event.

As apparent from Figure~\ref{Fig:offsets_mean_anomaly} (top panel), we see following trends in the offsets:

\begin{enumerate}
\item In L0, L1 and L2 lunations, the primary and secondary event offsets are systematically shifted relative to each other, which indicates an eccentric orbit.

\item The magnitude of the offsets varies in time and changes a sign periodically with a period of about 50 days, which indicates a presence of apsidal precession with this period.

\item There are no systematic shifts of the times of primary vs. secondary events in the data from the L3-L5 lunations,
which indicates that the eccentricity dropped to a value close to zero or entered chaotic evolution at a certain time between the L2 and L3 lunation,
i.e. about $72 \pm 6$ days after the impact.

\item The gradual trend of the offsets (both of the primary and the secondary events) during L0 lunation suggests a decrease of the orbital period during L0 or between L0 and L1.
% DONE: Harrison: I don’t think secular is the right word here? In my mind, a secular change is something that happens over very long timescales (i.e., tidal evolution, secular resonances, etc)
\end{enumerate}

In order to describe the above-mentioned features, we augmented the model in following way.
The orbit was assumed to be eccentric (with eccentricity $e_{0-2}$ and with apsidal precession with the rate ${\rm d}\varpi / {\rm d} t$) during the L0, L1 and L2 lunations,
while it was assumed to be circular during L3, L4 and L5 lunations.

So that to directly compare the orbital period of the circular orbit with the orbital period of the precessing eccentric orbit,
we redefine the orbital period as $P_{\rm orb}^{\rm post} = 360^{\circ}/(n + {\rm d}\varpi / {\rm d} t)$, which has now the meaning of a mean orbital period in the inertial frame (\citealp{MeyerEtAl2023}, use a term "stroboscopic orbit period").
% DONE: Alex: This is the 'stroboscopic orbit period' from Meyer et al 2021 and 2023.

To describe the apparent change of the mean orbital period during a few weeks after the impact, we used following two approaches.

{\it (a) Angular acceleration}

We added an angular acceleration term (see below for its physical reasoning) to the time variable mean anomaly

\begin{equation}
M (t) = M (t_{\rm imp}) + n (t-t_{\rm imp}) + A_c \exp(-(t-t_{\rm imp})/\tau),  \label{AngAccEq}
\end{equation}

assuming that

$$
n= \frac{360^{\circ}}{P_{\rm orb}^{\rm post}} - \frac{{\rm d}\varpi}{{\rm d} t},
$$
\begin{equation}
\dot{n} = A \exp(-(t-t_{\rm imp})/\tau)
\end{equation}

and $A_c = A \tau^2$. The epoch $t_{\rm imp}$ is the time of the impact.

%We stepped ${\rm d}\varpi / {\rm d} t$ and $\omega$ on a grid of 0..25.0~deg/day $\times$ 0..360~deg,
We stepped ${\rm d}\varpi / {\rm d} t$ and $\omega$ on a grid from 0 to 25.0~deg/day and from 0 to 360~deg, respectively,
with $M (t_{\rm imp})$, $A_c$, $\tau$, $e_{0-2}$, and $P_{\rm orb}^{\rm post}$ varied to get the best fit at each step.
Using the best-fit solution obtained with this augmented model, we constructed bottom panel of Figure~\ref{Fig:offsets_mean_anomaly}, which is analogous to its top panel,
but showing the offsets of the observed events with respect to the new improved solution (red points in the figure).

{\it (b) Orbital period modeled with a step function}

Besides the angular acceleration model described above, we also tested a more simple model to quantify the evolution of the mean orbital period following the impact, by modeling the period with step function.
We define $P_{\rm orb}^0$, a mean orbital period during an interval from the impact to the time $t_{\rm imp} + \Delta T$, when the orbital period drops from $P_{\rm orb}^0$ to $P_{\rm orb}^{\rm post}$ and remains constant since then.

Instead of Eq.~\ref{AngAccEq}, the evolution of the mean anomaly is then described as
\begin{eqnarray}
{\rm If\ } t - t_{\rm imp} < \Delta T &:& M (t) = M (t_{\rm imp}) + \left( \frac{360^{\circ}}{P_{\rm orb}^0} - \frac{{\rm d}\varpi}{{\rm d} t} \right) (t - t_{\rm imp}).\\
{\rm If\ } t - t_{\rm imp} \geq \Delta T &:& M (t) = M (t_{\rm imp}) + \left( \frac{360^{\circ}}{P_{\rm orb}^{\rm post}} - \frac{{\rm d}\varpi}{{\rm d} t} \right) (t - t_{\rm imp}) + \Delta M,
\end{eqnarray}

where $\Delta M = \frac{360^{\circ} \Delta T}{P_{\rm orb}^0} - \frac{360^{\circ} \Delta T}{P_{\rm orb}^{\rm post}}$ secures the continuity of $M$ at $t - t_{\rm imp} = \Delta T$.

%We also allowed a step-change of the eccentricity fifteen days after the impact, i.e., described the eccentricity with two-step function in the same way as in the approach (a).

We then fitted $P_{\rm orb}^0$, $\Delta T$, $P_{\rm orb}^{\rm post}$, and the other parameters of the system described above (except for $A_c$ and $\tau$, this time) to the data. 
In the same way as above we have calculated the mean anomaly offsets with respect to the best-fit solution obtained with this second approach,
and plotted them as green points in the bottom panel of Figure~\ref{Fig:offsets_mean_anomaly}.

The parameters of the best-fit solutions using the two approaches are listed in Table~\ref{tablePostImpact}. 
We obtained the same values for all the parameters of both solutions, except that they differ in introducing $A_c$ and $\tau$ or $P_{\rm orb}^0$ and $\Delta T$.
As the solution with the orbital period modeled as step function gives the same RMS residual as the solution with model of the angular acceleration of the mean anomaly (both give 0.0145 mag.),
and both solutions have very similar event timing offsets (see Figure~\ref{Fig:offsets_mean_anomaly})
we conclude that the two alternative models for the orbital period change are indistinguishable based on the quality of the fits to the data.

The offsets plotted in the bottom panel of Figure~\ref{Fig:offsets_mean_anomaly} show that with the augmented models we were able to describe the event timings in the L0, L1 and L2 lunations much better than using the simple circular orbit.
However, there is still present a systematic shift of the primary vs. secondary events in the data from the L0 lunation.
We therefore tried to further improve the fit in the following way.
We allowed for a step change of the eccentricity fifteen days after the impact (i.e., during the time interval between the L0 and L1 lunation).
The improved model describes the eccentricity as a two-step function with values of $e_0$ during the L0, $e_{1,2}$ during L1 and L2 and 0 during L3, L4 and L5 (with the former two values fitted as free parameters).
We obtained $e_0 = 0.042 \pm 0.021$, describing the event timings in L0 even better, but there still remained small systematic effects during L0 which we cannot explain.
Because it is based only on the data from the L0 lunation, the value of $e_0$ is also poorly constrained.
This experiment may indicate, that the dynamics of the orbit during the L0 lunation was more complex than what our simple model is able to describe. However, given the large uncertainties in the L0 lunation offsets, it is also possible that this is just a statistical fluke.
Given the above reasons, we do not adopt this model with the additional step in eccentricity as our final solution.

We also computed the 68th percentile of the absolute values of the offsets to quantify the fit in the sense of the event timings.
For the L1 and L2 lunations together, which have the smallest offsets, the 68th percentile is 0.7~deg., corresponding to 1.3~minutes in time.
For the entire post-impact apparition, the 68th percentile is 1.5~deg., corresponding to 2.8~minutes in time.
These values are estimates for $1\sigma$ rms residuals of the event times fitted with our model.

Examples of the long-period component data together with the synthetic
lightcurve of the best-fit solution are presented in Figures~\ref{Didymos_synth_postimpact} and~\ref{Didymos_synth_postimpactB}.

\begin{figure}[ht!]
\epsscale{0.75}  % The automatic scaling may be overridden with the command \epsscale{num}, where num is a scaling factor in decimal units, e.g., 0.80.
\plotone{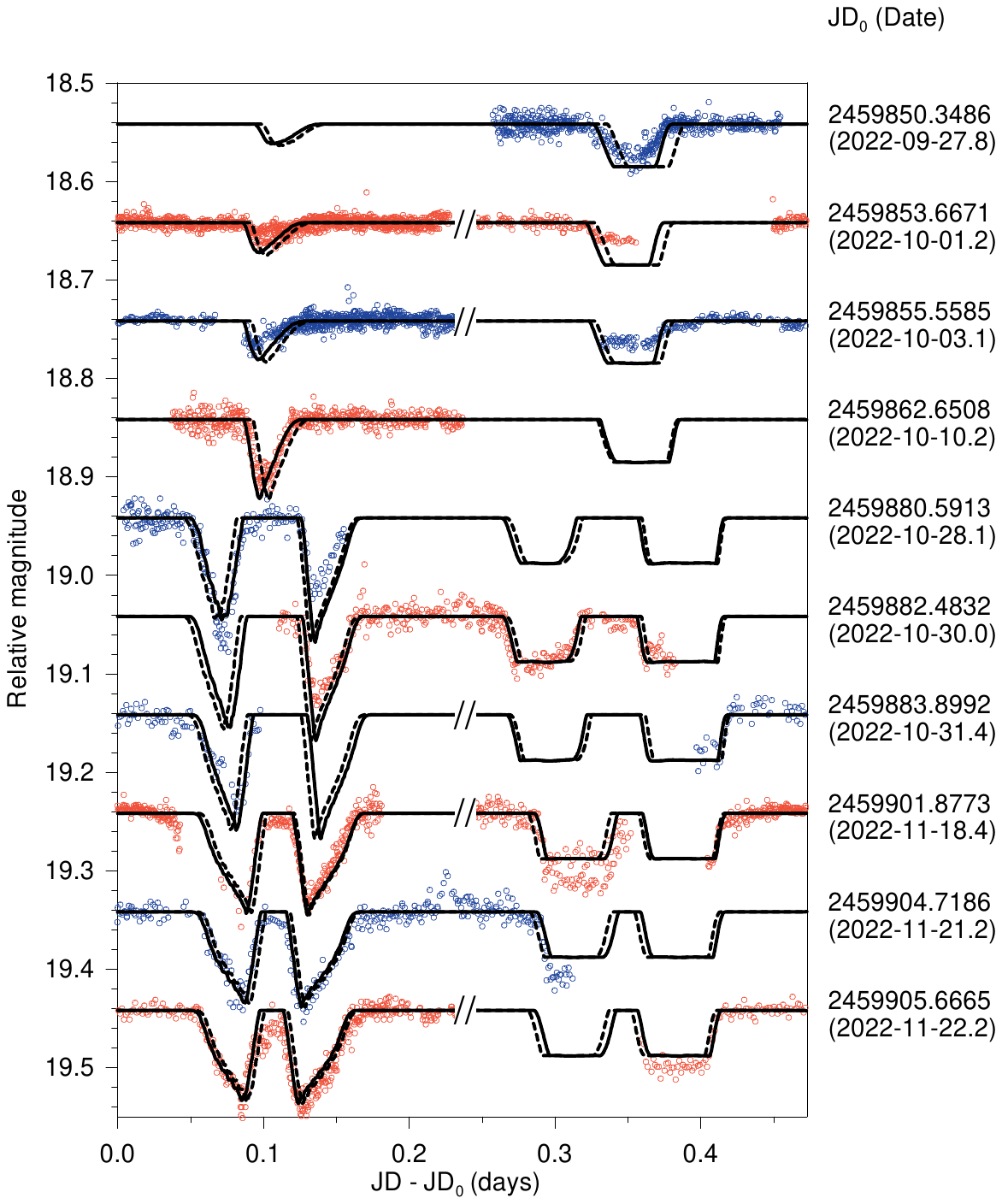}
\caption{Selected post-impact events from the orbital lightcurve component of the Didymos system.
The observed data are marked as points. (To avoid confusion between different data sets, different colors are used alternating on each plot.)
The solid curve represents the synthetic lightcurve for the best-fit solution with the angular acceleration.
The synthetic lightcurve for the best-fit solution with orbital period modeled as step function is identical within the thickness of the line used,
and is therefore not shown in the figure.
For comparison, the dashed curve is the model with the circular orbit and constant mean orbital period.
The primary and secondary events (the terms refer to which of the two bodies is occulted or eclipsed) are always shown on the left and right sides of the plots, respectively.
In some cases, the observations of a secondary event preceded that of a primary event (i.e., their order in the dataset is inverse of that shown on the plot).
In order to save space in the plot, we present these events in reverse order to how they were observed.
They are separated by the "//" symbol in the plot and one orbital period (0.496~d) is to be subtracted from the x coordinate of data points to the right from this separator.
\label{Didymos_synth_postimpact}}
\end{figure}

\begin{figure}[ht!]
\epsscale{0.75}  % The automatic scaling may be overridden with the command \epsscale{num}, where num is a scaling factor in decimal units, e.g., 0.80.
\plotone{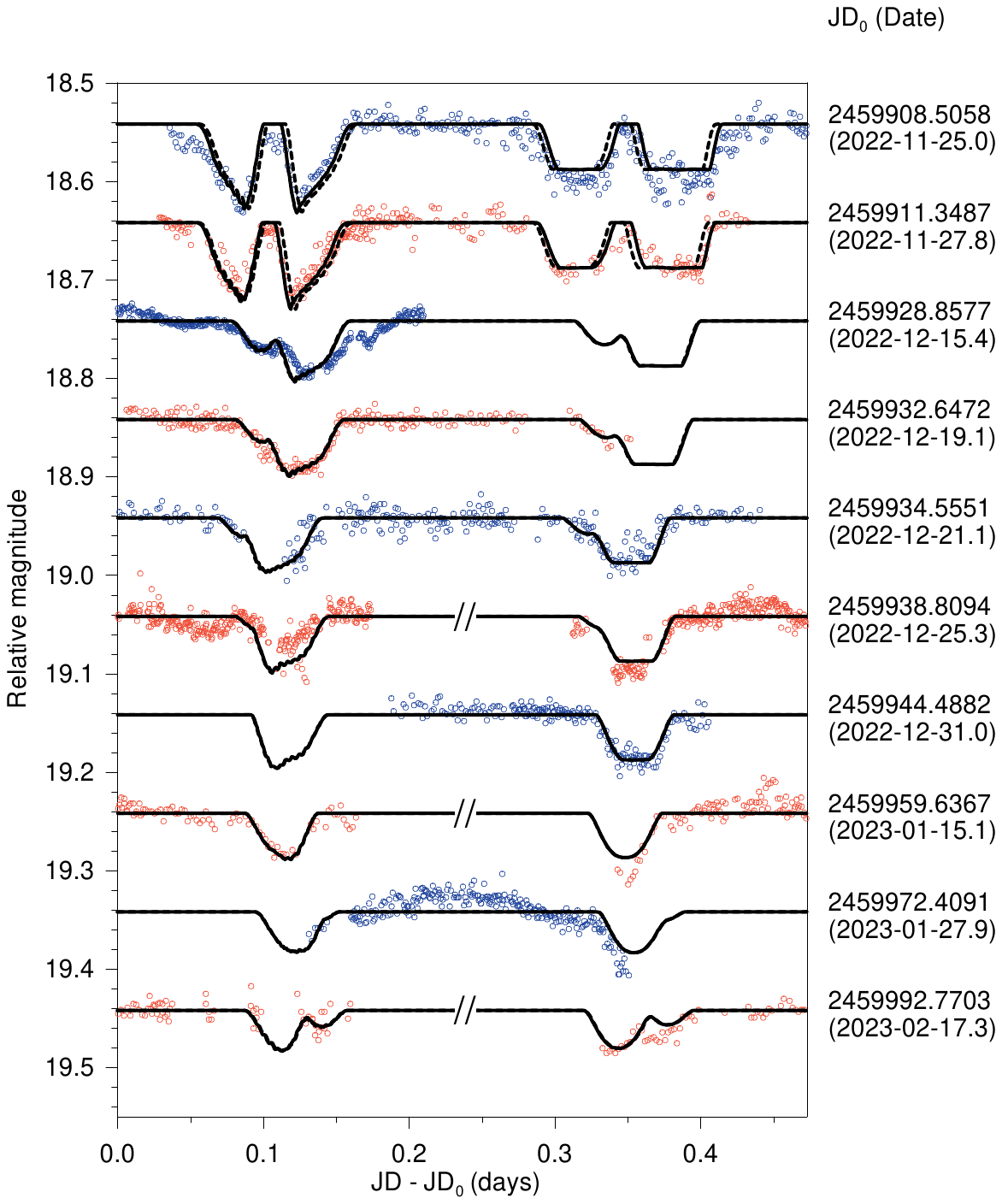}
\caption{Continuation of Figure~\ref{Didymos_synth_postimpact}.
\label{Didymos_synth_postimpactB}}
\end{figure}

%From $P_{\rm orb}^0$, $P_{\rm orb}^{\rm post}$, and the pre-impact orbital period $P_{\rm orb}^{\rm pre}$,
%we obtain that the period change
%that occured during the first $\sim$ 2 weeks
% a few days or weeks after the impact was $-32.77 \pm 0.30$~min., which changed to $-33.240 \pm 0.072$~min $14.0^{+12.5}_{-5.5}$~days after the impact (3$\sigma$).
% DONE: Cristina: What time period is included in "immediately after impact"? I assume this is some subset of the observations? The first few? all of L0? I assume this relates back to what you wrote on lines 208-209. Would there be a way to comment on if the period change is during L0? Or between L0 and L1? Was the period stable after this change?

Using the solution obtained with the model augmented with the angular acceleration, we also computed $P_{\rm orb}^{\rm imp}$ -- an orbital period at $t=t_{\rm imp}$, i.e., at the time immediately after the impact, from the time derivative of the mean anomaly.
We give its value in Table~\ref{tablePostImpact} as well.

From $P_{\rm orb}^{\rm post}$ and the pre-impact orbital period $P_{\rm orb}^{\rm pre}$,
we obtain that total Dimorphos' period change was $-33.240 \pm 0.072$~min.
The higher values of $P_{\rm orb}^{\rm imp}$ and $P_{\rm orb}^0$ than $P_{\rm orb}^{\rm post}$ indicate that the period change did not remain at a constant value, but it was lower (in absolute values)
shortly after the impact and it took a few weeks ($\Delta T = 15.5^{+12.0}_{-7.0}$ days, or $\tau = 14.3 \pm 6.5$ days) for the orbital period to stabilize.
From the solution with the exponentially decreasing angular acceleration
we obtain that the period change (caused by the DART impact) immediately after the impact was $P_{\rm orb}^{\rm imp}-P_{\rm orb}^{\rm pre} = -32.78^{+0.96}_{-0.31}$~min.
and the later period drop (caused by the following evolution of the system, see below) was $P_{\rm orb}^{\rm post}-P_{\rm orb}^{\rm imp} = -27^{+19}_{-58}$ seconds.
The solution with the period modeled as step function
gives the period change $P_{\rm orb}^0-P_{\rm orb}^{\rm pre} = -32.92 \pm 0.30$~min. and the period drop $P_{\rm orb}^{\rm post}-P_{\rm orb}^0 = -19 \pm 18$ seconds (all uncertainties are 3$\sigma$).
We note that $P_{\rm orb}^0$ represents only the lower limit of the orbital period just after the impact (it is the mean value over the time interval from $t_{\rm imp}$ to $t_{\rm imp}+\Delta T$).

One possible explanation for the small drop in the orbit period during the first few weeks following the DART impact is the idea of binary hardening due to the gravitational interaction of the ejecta cloud with
Didymos and Dimorphos \citep{RichardsonEtAl2023}. The DART impact released a significant amount of ejecta from Dimorphos,
from micrometer-sized dust to meter-sized boulders (e.g., \citealp{MorenoEtAl2023}, \citealp{RothEtAl2023}, \citealp{JewittEtAl2023}).
This observed ejecta is already outside Didymos's Hill sphere on escape trajectories.
However, a significant quantity of ejecta was likely initially bound to the system, where it would have made repeated close passes by Didymos and/or Dimorphos.
This material would have then been reaccreted or scattered out of the system during close approaches.
The net effect of this process could lead to an overall reduction in the orbit period, as Dimorphos would lose angular momentum by ejecting the bound debris. As discussed by \citet{RichardsonEtAl2023}, this orbital hardening will decay exponentially as the debris is cleared out of the system. In the approach involving the angular acceleration (see above), we included this effect as the exponentially decaying drag-like term in Equation~\ref{AngAccEq}.

However, this process will depend strongly on the mass-velocity distribution of ejecta at low speeds ($6 {\rm \ cm/s} < v < 25 {\rm \ cm/s}$), which is not well constrained.
Further investigation of this apparent period drop is needed.

Bottom panel of Figure~\ref{Fig:offsets_mean_anomaly} reveals that while most of the offsets of the observed events with respect to the two solutions are close to zero in L1 and L2 lunations,
indicating that the model explains the data well,
in lunations L3 to L5 (i.e., starting approximately 70 days after the DART impact) there occur larger random-like deviations of up to about 5 deg. (corresponding to $\sim$~10 minutes in time).
Since these deviations seem to have the same sign for both primary and secondary events (where both types of events occur close in time to each other), we cannot attribute them to eccentricity.
However, the deviations of mutual event timings may indicate variations in the orbital period.
\citet{MeyerEtAl2023} (also \citealp{MeyerEtAl2021}) showed that when Dimorphos enters a tumbling state, the variations of the orbital period,
driven by an exchange of angular momentum between the secondary's rotation and its orbit, become chaotic, varying on a time scale of twelve hours and can reach up to several minutes.
% DONE: Alex: variations driven by an exchange of angular momentum between the secondary and the orbit.
% DONE: Alex: also Meyer et al (2021).
Together with the suggested decrease in orbital eccentricity, these observed variations also suggest an onset of the tumbling state about 70 days after the impact.

\section{Mutual events ephemeris for 2022--2025}\label{Prediction}

In order to facilitate planning ground-based observations
in the next years, or to analyze the observations made in the pre- or post-impact apparitions, we computed times of mutual
events that occurred and will occur from July 2022 to the end of 2025. It is included here as a supplemental data file. [supplementary\_data.tar]

%It is available at\\
%\url{https://asu.cas.cz/~asteroid/Didymos_2022-%2025_events.txt}.

The prediction for pre- and post-impact apparition is made using the nominal solutions presented in Sections~\ref{PreImpactParams} and \ref{PostImpactParams}, respectively.
Due to the random-like deviations in the mutual events timings with respect to the nominal solution, which we see since approximately 70 days after the DART impact (see bottom panel of Figure~\ref{Fig:offsets_mean_anomaly}),
the uncertainties may be up to $\pm 10$~minutes larger for dates after 2022-12-02 than indicated in the link above.
A future evolution of the mutual orbit may increase the uncertainties further.

\section{Discussion and conclusions}\label{Sect:Discussion}

We have presented the results of modeling the mutual orbit of Didymos-Dimorphos system
using the mutual events data that were obtained within the photometric campaign carried out
in support of the DART mission.
We updated the parameters of the pre-impact orbit and obtained the solution for the new, post-impact orbit.

We obtained that the total Dimorphos' period change, i.e., the difference between the post-impact and the pre-impact orbital period, was $-33.240 \pm 0.072$~min. (all uncertainties are 3$\sigma$, unless otherwise stated).
The results show that while the majority of the period change ($\sim -33$~min.) occurred immediately after the impact, it was slightly smaller (in absolute values)
shortly after the impact and it took a few weeks for the orbital period to stabilize, i.e., it was followed by an additional change of a few tens of seconds.
By introducing two alternative models to describe the evolution of the orbital period in the days and weeks following the impact,
we found that the orbital period immediately after the impact was longer by $27^{+58}_{-19}$ seconds
(with the model of exponentially decreasing angular acceleration) or by $19 \pm 18$ seconds (with the model of the orbital period as step function) than its final value.

As the primary and secondary events were phase-shifted with respect to each other in the lunations L0 to L2 after the impact, and such shifts were not apparent in the lunations L3 to L5
(i.e., from about 70 days after the impact), we modeled the orbit as eccentric (with apsidal precession) during the L0, L1 and L2 lunations, while we modeled it as circular
during the L3, L4 and L5.
We obtain the eccentricity in the L0 to L2 lunations was $0.028 \pm 0.016$ and the apsidal precession rate of $7.3 \pm 2.0$~deg/day.
We also tried to fit the whole post-impact dataset with constant eccentricity, but we obtained an unsatisfactory fit.

The observed value of the apsidal precession rate is consistent with \citet{MeyerEtAl2021},
who predicted the precesion period to be from 22 to 62 days for Dimorphos' axis ratio $A_2/B_2$ from 1.4 to 1.3, respectively.
As the pre-impact value of $A_2/B_2$ was close to 1, see \citet{DalyEtAl2023b}, the observed apsidal precession rate suggests some reshaping of Dimorphos, as shown also by \citet{RichardsonEtAl2023}.

A gravitational oblateness parameter of Didymos, $J_2$, is estimated by \citet{NaiduEtAl2023} to be $0.090 \pm 0.008$, which is consistent with the oblate spheroidal shape of Didymos, assuming its uniform density.
Classical perturbation theory \citep{MurrayAndDermott2000} which assumes the secondary to be a point mass, gives an apsidal precession rate of around 13~deg/day, i.e., about twice our estimated rate.
However, as shown by \citet{CukAndNesvorny2010} (see also \citealp{MeyerEtAl2023}), the apsidal precession might be dominated by the perturbations from the prolate and librating secondary, rather than the oblate primary.

Since approximately 70 days after the DART impact, we see random-like deviations of up to $\sim$~10 minutes in the mutual events timings with respect to the nominal solution, which we cannot attribute to the eccentricity.
Those deviations may indicate short-term variations in the orbital period.
The observed variations together with the suspected decrease of the orbital eccentricity at about the same time suggest that Dimorphos entered a tumbling state around 70 days after the impact.
Due to the conservation of angular momentum, the binary eccentricity can only change as a result of a change in orbital angular momentum. If the post-impact Didymos binary is considered a closed system, then the only sources of angular momentum for the mutual orbit are in the primary and secondary rotations. The primary’s spin/shape has not changed by a measurable amount \citep{DurechAndPravec2023}, which suggests that the secondary would have had to undergo some change in angular momentum to account for the eccentricity change. A natural way to do this is if the secondary enters an excited rotation state (i.e., tumbling) where angular momentum can be transferred from the mutual orbit to the secondary’s rotation. We consider this as the most plausible explanation for the eccentricity change.

From the depths of the mutual events, we estimated the pre-impact Dimorphos-Didymos size ratios was $D_{2,V}/D_{1,V} = 0.223 \pm 0.004$ and the post-impact is $0.202 \pm 0.006$ (1$\sigma$ uncertainties).
The difference between the two values indicates the reduction of Dimorphos' volume by ($9 \pm 3) \%$ (1$\sigma$) as the result of the impact.
This reduction is about an order of magnitude larger than the estimate for the mass of the ejecta from observations and modeling (the largest estimate of $5.5\times10^7$~kg is by \citealp{FerrariEtAl2023}; see \citealp{RichardsonEtAl2023} for a list of published estimates) and therefore it deserves further discussion. First, we point out the relatively large uncertainty of our estimate, so the inferred reduction of Dimorphos's volume is only a marginally significant detection.  As the DART impact caused global deformation and resurfacing of Dimorphos (see \citealp{RaducanEtAl2023}), the reduction of the Dimorphos volume might suggest an impact-induced change in its macroporosity.
Further, the Dimorphos-Didymos size ratios are based on depths of total secondary events and an assumption that albedos of both components were the same, both before and after the impact. Albedo differences can introduce a systematic error into the estimates.

\begin{figure}[ht!]
\epsscale{1.4}  % The automatic scaling may be overridden with the command \epsscale{num}, where num is a scaling factor in decimal units, e.g., 0.80.
\plotone{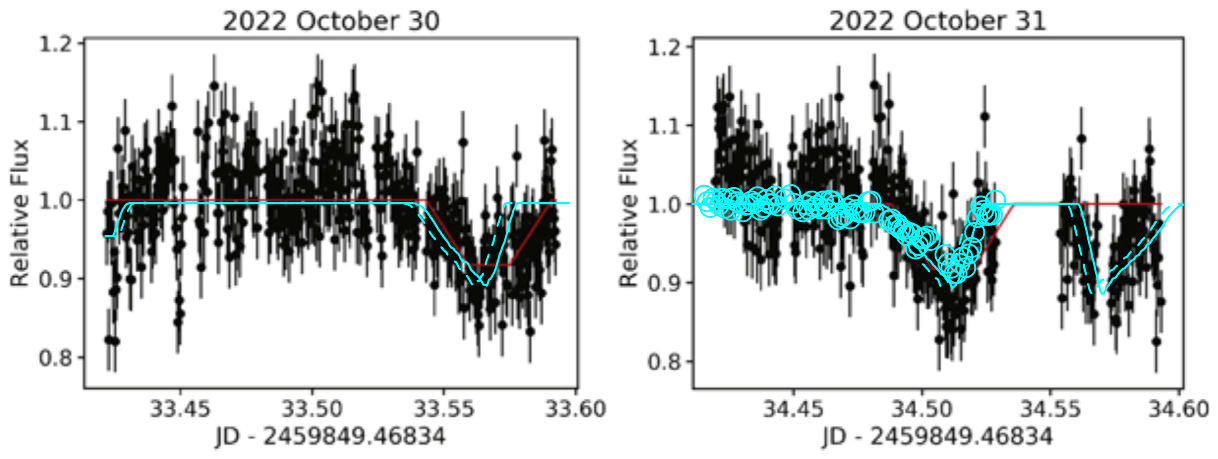}
\caption{Figure~1 from \citet{GudebskiEtAl2023} with the orbital lightcurve component (cyan circles), the synthetic lightcurve for the best-fit solution with the angular acceleration (solid cyan curve) and
the model with the circular orbit and constant mean orbital period (dashed cyan curve) from this work overplotted.
\label{Fig:Gudebski}}
\end{figure}

The change of the orbital period was also estimated by \citet{GudebskiEtAl2023}, who compared the mutual events prediction from SP22 with events observed on 2022 October 30 and 31, and obtained $\Delta P= -34.2 \pm 0.1$~min.  
To compare their results with ours, we constructed
Figure~\ref{Fig:Gudebski}, which contains the bottom part of their Figure~1, to which we had added the data used in our analysis and the synthetic curves of our nominal solutions shown in Figure~\ref{Didymos_synth_postimpact}. The figure shows that the lightcurve decomposition and the times of the mutual events derived by  Gudebski et al. are fairly consistent with our nominal solution (although they did not identify the second part of the primary event from October 31 in their data). However, the description of the method they used to derive the new orbital period from the data (especially, the first paragraph of their Section~3) is quite vague and a description of how they estimated its uncertainty is also missing. They also did not take into account the uncertainty of the event ephemeris from SP22, which was $\pm 20$~minutes (3$\sigma$) at the end of October 2022. Therefore, we are unable to compare their result with the conclusions of this paper.

In this work, we used a simplified model, that did not include a full two-body interaction and describes the orbit parameters as dynamically independent variables.
The advantage of this approach is that it does not involve any a priori dynamical assumptions and therefore serves as an independent check of more complex models.
An independent derivation of the post-impact mutual orbit based on an analysis of mutual event timings by integrating Dimorphos' orbit using a 2-dimensional model has been made by \citet{NaiduEtAl2023}
Their estimated parameters are in agreement with our values within the uncertainties.

\begin{acknowledgments}
The work by P.P. and P.S. was supported by the Grant Agency of the Czech Republic, grant 20-04431S. This work was supported in part by the DART mission, NASA Contract \#80MSFC20D0004 to JHU/APL. This work has been supported by the French government, through the UCA J.E.D.I. Investments in the Future project managed by the National Research Agency (ANR) with the reference number ANR-15-IDEX-01. Access to computing and storage facilities owned by parties and projects contributing to 
the National Grid Infrastructure Meta-Centrum provided under the program “Projects of Large Research, Development, and Innovations Infrastructures” (CESNET LM2015042), and the CERIT Scientific Cloud LM2015085, is greatly appreciated. The work of S.P.N. and S.R.C. was carried out at the Jet Propulsion Laboratory, California Institute of Technology, under a contract with the National Aeronautics and Space Administration (No. 80NM0018D0004). We thank two anonymous reviewers for their valuable comments that led us to improve the paper substantially. 
\end{acknowledgments}

\bibliography{sajri-refs}{}
\bibliographystyle{aasjournal}

\end{document}